\documentclass[conference]{IEEEtran}

\usepackage[utf8]{inputenc}
\usepackage[T1]{fontenc}
\usepackage{amsmath}
\usepackage{amssymb}
\usepackage{tabularx}
\usepackage{tikz}
\usetikzlibrary{arrows}
\usetikzlibrary{positioning}
\usetikzlibrary{shadows}
\usepackage{subfigure}
\usepackage{tikz-er2}
\usepackage{pifont}
\usepackage{amsthm}
\usepackage{epsfig, graphics, graphicx, subfigure}
\usepackage{moreverb}
\usepackage{fancyhdr}
\usepackage{color}
\usepackage{rotating}
\usepackage{longtable}
\usepackage{pifont}
\usepackage{capt-of}
\usepackage{listings}
\usepackage{stmaryrd}
\usepackage{pstricks}
\usepackage{rotating}
\usepackage{multirow}
\usepackage{tabularx}

\renewcommand{\baselinestretch}{.935}

\newtheorem{example}{Example}

\DeclareMathOperator{\Vp}{V^\dagger}
\DeclareMathOperator{\V}{V}
\newcommand{\dirac}[1]{\ensuremath{|#1 \rangle}}

\title{Clarification on the Mapping\\of Reversible Circuits to the NCV-$|v_1\rangle$-Library}
\author{Zahra Sasanian \and Robert Wille \and D. Michael Miller}

\author{Zahra Sasanian$^1$, Robert Wille$^{2}$,  D. Michael Miller$^1$\\[1em]
$^1$Department of Computer Science, University of Victoria, Canada\\
$^2$Group for Computer Architecture, University of Bremen, Germany\\
sasanian@uvic.ca\hspace{1cm}rwille@informatik.uni-bremen.de\hspace{1cm}mmiller@uvic.ca}

\begin{document}
\maketitle

\section{Introduction}

In~\cite{SWM:2012} and motivated by the theoretical discussion on physical realizations from~\cite{MS:00}, a new quantum gate library (the NCV-$|v_1\rangle$ library) for electronic design automation of quantum circuits has been proposed. Here, \emph{qudits} instead of qubits are assumed, i.e.~a basic building block which does not rely on 
a two level quantum system but a (multiple-valued) \mbox{$d$-level} quantum system is assumed.
However, the descriptions in~\cite{SWM:2012} on the foundation of this new library remained brief.
This technical report provides an extended description of the applied ideas and concepts.

\section{Reversible Gates \& Circuits}

A logic function $f:\mathbb{B}^n\rightarrow \mathbb{B}^m$ over inputs \mbox{$X=\{x_1,\dots , x_n\}$}
is \emph{reversible}
if and only if
\begin{itemize}
\item its number of inputs is equal to its number of outputs (i.e.~$n=m$) and
\item it maps each input pattern to a unique output pattern.
\end{itemize}
Otherwise, the function is termed \emph{irreversible}.
In other words, a reversible function represents a bijection.

A reversible function can be realized by a circuit \mbox{$G=g_1g_2\dots g_d$} comprised of a cascade of reversible gates $g_i$, where~$d$ is the number of gates. Fanouts and feedback are not directly allowed~\cite{NC:2000}. Several different reversible gates have been introduced including the Toffoli gate~\cite{Tof:1980}, the Fredkin gate~\cite{FT:82b}, and the Peres gate~\cite{Per:85}.
In the following, we focus on Toffoli gates which are universal gates, i.e.~all reversible functions can be realized by means of this gate type alone~\cite{Tof:1980}.

A {\em multiple control Toffoli gate\/} has a \emph{target line} $x_j$ and \emph{control lines} $\{x_{i_1},x_{i_2},\ldots ,x_{i_k}\}$. This gate maps $(x_1, x_2, \ldots, x_j, \ldots, x_n)$ to $(x_1,x_2,\ldots, x_{i_1} x_{i_2}\ldots x_{i_k} \oplus x_j, \ldots, x_n)$. That is, the target line is inverted if all control lines are set to~1; otherwise the value of the target line is passed through unchanged.
A Toffoli gate with no control lines always inverts the target line and is a \emph{NOT gate}.  A Toffoli gate with a single control line is called a \emph{controlled-NOT gate} (also known as the CNOT gate).  The case of two control lines is the original gate defined by Toffoli.  For brevity, we refer to a multiple-control Toffoli gate as a Toffoli gate.

In the following, a Toffoli gate is denoted by the tuple $T(C,t)$ where $C\subset X$ is the possibly empty set of control lines and $t\in X\setminus C$ is the target line.
Note that the control lines and unconnected lines pass through a gate unchanged.
For drawing circuits, we follow the established convention of using the symbol $\oplus$ to denote the target line 
and solid black circles to indicate control connections for the gate.

\begin{figure}[t]
\centering
\begin{tikzpicture}
\draw[line width=0.300000] (1.000000,1.750000) -- (5.000000,1.750000);
\draw (0.900000,1.750000) node [left] {$x_1=1$};
\draw (5.100000,1.750000) node [right] {$f_1=1$};
\draw[line width=0.300000] (1.000000,1.250000) -- (5.000000,1.250000);
\draw (0.900000,1.250000) node [left] {$x_2=1$};
\draw (5.100000,1.250000) node [right] {$f_2=0$};
\draw[line width=0.300000] (1.000000,0.750000) -- (5.000000,0.750000);
\draw (0.900000,0.750000) node [left] {$x_3=1$};
\draw (5.100000,0.750000) node [right] {$f_3=0$};
\draw[line width=0.300000] (1.000000,0.250000) -- (5.000000,0.250000);
\draw (0.900000,0.250000) node [left] {$x_4=1$};
\draw (5.100000,0.250000) node [right] {$f_4=0$};
\draw[line width=0.300000] (1.500000,0.250000) -- (1.500000,1.750000);
\draw[fill] (1.500000,1.750000) circle (0.100000);
\draw[fill] (1.500000,1.250000) circle (0.100000);
\draw[line width=0.3] (1.5,0.25) circle (0.2) (1.5,0.05) -- (1.5,0.45);
\draw[line width=0.300000] (2.500000,1.250000) -- (2.500000,1.750000);
\draw[fill] (2.500000,1.750000) circle (0.100000);
\draw[line width=0.3] (2.5,1.25) circle (0.2) (2.5,1.05) -- (2.5,1.45);
\draw[line width=0.300000] (3.500000,0.750000) -- (3.500000,1.750000);
\draw[fill] (3.500000,1.750000) circle (0.100000);
\draw[line width=0.3] (3.5,0.75) circle (0.2) (3.5,0.55) -- (3.5,0.95);
\draw[line width=0.300000] (4.500000,0.250000) -- (4.500000,1.750000);
\draw[fill] (4.500000,0.250000) circle (0.100000);
\draw[line width=0.3] (4.5,1.75) circle (0.2) (4.5,1.55) -- (4.5,1.95);

\node at (2,1.9) {$1$};
\node at (2,1.4) {$1$};
\node at (2,0.9) {$1$};
\node at (2,0.4) {$0$};

\node at (3,1.9) {$1$};
\node at (3,1.4) {$0$};
\node at (3,0.9) {$1$};
\node at (3,0.4) {$0$};

\node at (4,1.9) {$1$};
\node at (4,1.4) {$0$};
\node at (4,0.9) {$0$};
\node at (4,0.4) {$0$};
\end{tikzpicture}
\caption{Reversible circuit}
\label{fig:circ_example} 
\end{figure}
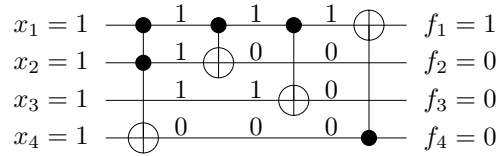

\begin{example}
Fig.~\ref{fig:circ_example} shows a reversible circuit composed of $n=4$ circuit lines and $d=4$~Toffoli gates.
This circuit maps e.g.~the input pattern $1111$ to the output pattern~$1000$ (as shown in Fig.~\ref{fig:circ_example}). Inherently, every computation can be performed in both directions (i.e.~computations towards the outputs \emph{and} towards the inputs can be performed).
\end{example}

\section{Quantum Gates \& Circuits}

The basic building block for a quantum computer is the qubit. A qubit is a two
level quantum system, described by a two dimensional complex Hilbert space. The
two orthogonal quantum states 
$|0 \rangle \equiv \left(^{1}_{0}\right)$ and
$|1 \rangle \equiv \left(^{0}_{1}\right)$
are used to represent the values 0 and 1. Any state of a qubit may be written as 
$
| \Psi \rangle = \alpha |0 \rangle + \beta |1 \rangle,
$
where $\alpha$ and $\beta$ are complex numbers with the following condition $|\alpha|^2 + |\beta|^2 = 1$. 
The quantum state of a single qubit is denoted by the vector 
${\alpha \choose \beta}$. 
The state of a quantum system with $n > 1$ qubits is given by an element of the
tensor product of the single state spaces and can be represented as a normalized
vector of length $2^n$, called the state vector. The state vector is changed
through multiplication of appropriate $2^n \times 2^n$ unitary matrices
\cite{NC:2000}.

Since this allows an infinite number of qubit-values and corresponding operations, 
researchers defined proper models and gate libraries in order to realize Boolean functions
in quantum logic. Here, the following two libraries are considered.

\subsection{NCV-Library}

\begin{figure}[t]
\centering
\begin{picture}(205,100)

\put(0,50){\large $0$}

\put(200,50){\large $1$}

\put(100,100){\large $v_1$}

\put(100,0){\large $v_0$}


\put(12,65){\vector(2,1){85}}
\put(50,92){\small $\Vp$} 
\put(97,101){\vector(-2,-1){85}}
\put(50,70){\small $\V$} 

\put(115,107){\vector(2,-1){85}}
\put(150,90){\small $\Vp$} 
\put(199,58){\vector(-2,1){85}}
\put(150,70){\small $\V$}


\put(114,8){\vector(2,1){85}}
\put(150,31){\small $\V$} 
\put(97,0){\vector(-2,1){85}}
\put(50,10){\small $\Vp$}

\put(198,43){\vector(-2,-1){83}} 
\put(150,10){\small $\Vp$} 
\put(12,50){\vector(2,-1){85}}
\put(50,31){\small $\V$}

\put(15,54){\vector(1,0){180}}
\put(195,54){\vector(-1,0){180}}

\put(107,57){\small NOT} 
\put(106,15){\vector(0,1){81}}
\put(106,96){\vector(0,-1){81}}

\end{picture}
\caption{State transitions for NOT, CNOT, $\V$, and $\Vp$ operations}\label{fig:not_v_v+_transitions}
\end{figure}
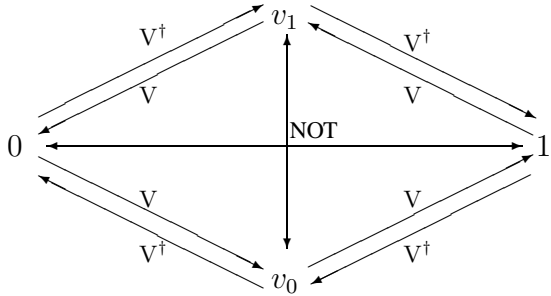

The quantum gate library introduced by Barenco et al.~\cite{BBC+:95} is mostly applied 
in the development of design methods for quantum circuits. 
Here, the following set of quantum gates is considered:
\begin{itemize}
\item NOT gate $T(\emptyset,t)$: A single qubit~$t$ is inverted. 
\item Controlled NOT (CNOT) gate $T(\{c\},t)$: The target qubit~$t$ is inverted if the control qubit~$c$ is 1.
\item Controlled $\V$ gate $\V(\{c\},t)$: A $\V$ operation is performed on the target qubit~$t$ if the control qubit~$c$ is 1. 
The $\V$ operation is also known as the square root of
NOT, since two consecutive $\V$ operations are equivalent to an inversion.  
\item Controlled $\Vp$ gate $\Vp(\{c\},t)$: A $\Vp$ operation is performed on the target qubit~$t$ if the control qubit~$c$ is 1. The $\Vp$ gate performs the inverse operation of the $\V$ gate,
i.e.~$\Vp = \V^{-1}$.
\end{itemize}
More precisely, these gates transform the target qubit~$t$ as specified by the unitary matrices
\begin{align*}
  NOT = \left(\begin{smallmatrix} 0 & 1 \\ 1 & 0 \end{smallmatrix}\right), 
  \V  = \tfrac{1+i}{2}\left(\begin{smallmatrix} 1 & -i \\ -i & 1 \end{smallmatrix}\right)
  \text{, and }
  \Vp = \tfrac{1-i}{2}\left(\begin{smallmatrix} 1 & i \\ i & 1 \end{smallmatrix}\right).
\end{align*}
If the input signals and all control lines are restricted to
Boolean values only, a 4-valued logic results where the value of each qubit is restricted to one of $\{ 0, 1, v_0, v_1 \}$ with
$ v_0 = \frac{1+i}{2} {1 \choose -i}$ and
$ v_1 = \frac{1+i}{2} {-i \choose 1}$.
Fig.~\ref{fig:not_v_v+_transitions} shows the resulting transitions with respect to
the possible NOT, $\V$, and $\Vp$ operations.
This is sufficient to realize every reversible function  as a quantum
circuit~\cite{BBC+:95}. 
Furthermore, this keeps the gate library simple enough to become physically realizable.
In the following, this model is called \emph{NCV gate library}.

\begin{figure}[t]
\centering
\begin{tikzpicture}
\draw[line width=0.300000] (1.000000,1.750000) -- (7.000000,1.750000);
\draw (0.900000,1.750000) node [left] {$x_1=1$};
\draw (7.100000,1.750000) node [right] {$f_1=1$};
\draw[line width=0.300000] (1.000000,1.250000) -- (7.000000,1.250000);
\draw (0.900000,1.250000) node [left] {$x_2=1$};
\draw (7.100000,1.250000) node [right] {$f_2=0$};
\draw[line width=0.300000] (1.000000,0.750000) -- (7.000000,0.750000);
\draw (0.900000,0.750000) node [left] {$x_3=1$};
\draw (7.100000,0.750000) node [right] {$f_3=0$};
\draw[line width=0.300000] (1.000000,0.250000) -- (7.000000,0.250000);
\draw (0.900000,0.250000) node [left] {$x_4=1$};
\draw (7.100000,0.250000) node [right] {$f_4=0$};
\draw[line width=0.300000] (1.500000,0.250000) -- (1.500000,1.250000);
\draw[fill] (1.500000,1.250000) circle (0.100000);
\draw[line width=0.3,fill=white] (1.3,0.05) rectangle ++(0.4,0.4); \node at (1.5,0.25) {\sf $\V$};
\draw[line width=0.300000] (2.500000,0.250000) -- (2.500000,1.750000);
\draw[fill] (2.500000,1.750000) circle (0.100000);
\draw[line width=0.3,fill=white] (2.3,0.05) rectangle ++(0.4,0.4); \node at (2.5,0.25) {\sf $\V$};
\draw[line width=0.300000] (3.500000,1.250000) -- (3.500000,1.750000);
\draw[fill] (3.500000,1.750000) circle (0.100000);
\draw[line width=0.3] (3.5,1.25) circle (0.2) (3.5,1.05) -- (3.5,1.45);
\draw[line width=0.300000] (4.500000,0.250000) -- (4.500000,1.250000);
\draw[fill] (4.500000,1.250000) circle (0.100000);
\draw[line width=0.3,fill=white] (4.3,0.05) rectangle ++(0.4,0.4); \node at (4.5,0.25) {\sf $\Vp$};
\draw[line width=0.300000] (5.500000,0.750000) -- (5.500000,1.750000);
\draw[fill] (5.500000,1.750000) circle (0.100000);
\draw[line width=0.3] (5.5,0.75) circle (0.2) (5.5,0.55) -- (5.5,0.95);
\draw[line width=0.300000] (6.500000,0.250000) -- (6.500000,1.750000);
\draw[fill] (6.500000,0.250000) circle (0.100000);
\draw[line width=0.3] (6.5,1.75) circle (0.2) (6.5,1.55) -- (6.5,1.95);

\node at (2,1.9) {$1$};
\node at (2,1.4) {$1$};
\node at (2,0.9) {$1$};
\node at (2,0.4) {$v_1$};

\node at (3,1.9) {$1$};
\node at (3,1.4) {$1$};
\node at (3,0.9) {$1$};
\node at (3,0.4) {$0$};

\node at (4,1.9) {$1$};
\node at (4,1.4) {$0$};
\node at (4,0.9) {$1$};
\node at (4,0.4) {$0$};

\node at (5,1.9) {$1$};
\node at (5,1.4) {$0$};
\node at (5,0.9) {$1$};
\node at (5,0.4) {$0$};

\node at (6,1.9) {$1$};
\node at (6,1.4) {$0$};
\node at (6,0.9) {$0$};
\node at (6,0.4) {$0$};
\end{tikzpicture}
\caption{Quantum circuit using the NCV gate library}
\label{fig:qua_barenco}
\end{figure}
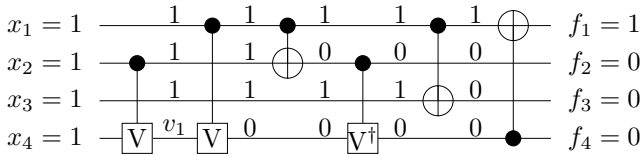

\begin{example}
Fig.~\ref{fig:qua_barenco} shows a quantum circuit composed of $n=4$ circuit lines and $d=6$~quantum gates.
This circuit again maps e.g.~the input pattern $1111$ to the output pattern~$1000$, but in contrast to the circuit from Fig.~\ref{fig:circ_example} quantum values and quantum operations are utilized for this purpose.
\end{example}

\subsection{NCV-$|v_1\rangle$-Library}
 
Although the Barenco gate model is universal, i.e.~every Boolean function can be 
represented by it~\cite{BBC+:95}, extensions of it have been introduced recently.
Here, we
additionally consider the quantum gate library introduced in~\cite{SWM:2012} which is based on the physical foundation from~\cite{MS:00}. Here, \emph{qudits} instead of qubits are assumed, i.e.~a basic building block which does not rely on 
a two level quantum system, but a (multiple-valued) $d$-level quantum system is assumed.
Any state of a qudit may be written as
$|\Psi\rangle = c_0 |0\rangle + c_1 |1\rangle + \ldots + c_{d-1} |d-1\rangle$
where $c_i$ for all $i=0,\ldots,d-1$ are complex numbers such that
$|c_0|^2+|c_1|^2 + \ldots + |c_{d-1}|^2 = 1$.
Similar to qubits, the states of a qudit are represented by a 
state vector.
The state vector is changed through multiplication of appropriate 
unitary matrices. In the case of an uncontrolled transformation the dimension of the
unitary matrices is $d \times d$, in the case of an controlled transformation the
dimension is $d^2 \times d^2$.
However, in contrast to a qubit, the controlled gates for qudits perform the respective
operation not when the control line is~\dirac{1}, but rather when the control line is set to the value~$\dirac{d-1}$.

In~\cite{MS:00}, a corresponding gate library for qudits has been presented and physically realized.
In~\cite{SWM:2012}, this model is adopted with a 4-level logic, i.e. $d=4$. The basic states in that order are $0$, $v_0$, $1$, and $v_1$. 
As already explained, the controlled gates are only transforming the target qudit if the value of the control line is set to $\dirac{d-1} \equiv v_1$.
We emphasize this fact by labeling the control connections for the respective gates with $v_1$.

The libary is composed of the three unitary gates (i.e.~gates without a control line) performing the NOT, $\V$, and $\Vp$ operation as well as single-control versions of these gates. More precisely, these gates transform the target qubit~$t$ as specified by the 
unitary matrices
\begin{align*}
  NOT = \left(\begin{smallmatrix}0&0&1&0\\ 0&0&0&1\\ 1&0&0&0\\ 0&1&0&0\\\end{smallmatrix}\right),
  \V = \left(\begin{smallmatrix}0&0&0&1\\ 1&0&0&0\\ 0&1&0&0\\ 0&0&1&0\\\end{smallmatrix}\right),
  \Vp = \left(\begin{smallmatrix}0&1&0&0\\ 0&0&1&0\\ 0&0&0&1\\ 1&0&0&0\\\end{smallmatrix}\right).
\end{align*}
Again, restricting the inputs to Boolean values still allows for the realization of any arbitrary reversible functions.
In the following, this model is called NCV-$|v_1\rangle$ gate library.

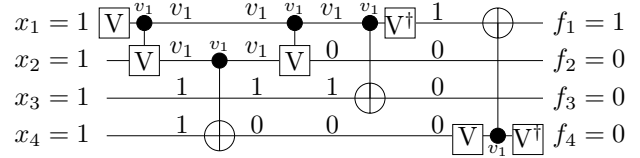
\begin{figure}[t]
\centering
\begin{tikzpicture}
\draw[line width=0.300000] (1.000000,1.750000) -- (6.800000,1.750000);
\draw (0.900000,1.750000) node [left] {$x_1=1$};
\draw (6.800000,1.750000) node [right] {$f_1=1$};
\draw[line width=0.300000] (1.000000,1.250000) -- (6.800000,1.250000);
\draw (0.900000,1.250000) node [left] {$x_2=1$};
\draw (6.800000,1.250000) node [right] {$f_2=0$};
\draw[line width=0.300000] (1.000000,0.750000) -- (6.800000,0.750000);
\draw (0.900000,0.750000) node [left] {$x_3=1$};
\draw (6.800000,0.750000) node [right] {$f_3=0$};
\draw[line width=0.300000] (1.000000,0.250000) -- (6.800000,0.250000);
\draw (0.900000,0.250000) node [left] {$x_4=1$};
\draw (6.800000,0.250000) node [right] {$f_4=0$};

\draw[line width=0.3,fill=white] (0.9,1.55) rectangle ++(0.4,0.4); \node at (1.1,1.75) {\sf $\V$};

\draw[line width=0.300000] (1.500000,1.250000) -- (1.500000,1.750000);
\draw[fill] (1.500000,1.750000) circle (0.100000);
\node[anchor=south] at (1.500000,1.750000) {\scriptsize $v_1$};
\draw[line width=0.3,fill=white] (1.3,1.05) rectangle ++(0.4,0.4); \node at (1.5,1.25) {\sf $\V$};
\draw[line width=0.300000] (2.500000,0.250000) -- (2.500000,1.250000);
\draw[fill] (2.500000,1.250000) circle (0.100000);
\node[anchor=south] at (2.500000,1.250000) {\scriptsize $v_1$};
\draw[line width=0.3] (2.5,0.25) circle (0.2) (2.5,0.05) -- (2.5,0.45);
\draw[line width=0.300000] (3.500000,1.250000) -- (3.500000,1.750000);
\draw[fill] (3.500000,1.750000) circle (0.100000);
\node[anchor=south] at (3.500000,1.750000) {\scriptsize $v_1$};
\draw[line width=0.3,fill=white] (3.3,1.05) rectangle ++(0.4,0.4); \node at (3.5,1.25) {\sf $\V$};
\draw[line width=0.300000] (4.500000,0.750000) -- (4.500000,1.750000);
\draw[fill] (4.500000,1.750000) circle (0.100000);
\node[anchor=south] at (4.500000,1.750000) {\scriptsize $v_1$};
\draw[line width=0.3] (4.5,0.75) circle (0.2) (4.5,0.55) -- (4.5,0.95);

\draw[line width=0.3,fill=white] (4.7,1.55) rectangle ++(0.4,0.4); \node at (4.9,1.75) {\sf $\Vp$};

\draw[line width=0.3,fill=white] (5.6,0.00) rectangle ++(0.4,0.4); \node at (5.8,0.25) {\sf $\V$};
\draw[line width=0.3,fill=white] (6.4,0.00) rectangle ++(0.4,0.4); \node at (6.6,0.25) {\sf $\Vp$};

\draw[line width=0.300000] (6.20000,0.250000) -- (6.200000,1.750000);
\draw[fill] (6.200000,0.250000) circle (0.100000);
\node[anchor=north] at (6.200000,0.250000) {\scriptsize $v_1$};
\draw[line width=0.3] (6.2,1.75) circle (0.2) (6.2,1.55) -- (6.2,1.95);

\node at (2,1.9) {$v_1$};
\node at (2,1.4) {$v_1$};
\node at (2,0.9) {$1$};
\node at (2,0.4) {$1$};

\node at (3,1.9) {$v_1$};
\node at (3,1.4) {$v_1$};
\node at (3,0.9) {$1$};
\node at (3,0.4) {$0$};

\node at (4,1.9) {$v_1$};
\node at (4,1.4) {$0$};
\node at (4,0.9) {$1$};
\node at (4,0.4) {$0$};

\node at (5.4,1.9) {$1$};
\node at (5.4,1.4) {$0$};
\node at (5.4,0.9) {$0$};
\node at (5.4,0.4) {$0$};
\end{tikzpicture}
\caption{Quantum circuit NCV-$|v_1\rangle$ gate library}
\label{fig:qua_new_lib}
\end{figure}

\begin{example}
Fig.~\ref{fig:qua_new_lib} shows a quantum circuit composed of $n=4$ circuit lines and $d=5$~quantum gates. This circuit performs the same computation as the circuits from Fig.~\ref{fig:circ_example} and Fig.~\ref{fig:qua_barenco}, but is composed of gates from the NCV-$|v_1\rangle$ introduced in~\cite{SWM:2012}.
\end{example}

\section{Mapping Reversible Circuits\\to Quantum Circuits}\label{sec:mapping}

Since any quantum operation can be represented by a unitary
matrix~\cite{NC:2000}, each quantum circuit is inherently reversible.
Consequently, every reversible circuit can be transformed to a quantum circuit.
Motivated by this, synthesis of Boolean components of quantum circuits is usually conducted 
in two steps: First, the desired logic is synthesized as reversible circuit.
Afterwards, each gate of the resulting circuit is mapped to a corresponding cascade of quantum gates.
To this end and depending on the addressed gate library, different mapping schemes have been proposed.

\subsection{Mapping to Gates from the NCV Library}\label{sec:mapping_ncv}

Mapping of reversible gates to NCV gates has intensely been considered in the past.
Originally, Barenco et al.~proposed the initial mappings in~\cite{BBC+:95}. Afterwards,
further improvements have been introduced e.g.~in~\cite{MYDM:2005} and, more recently, in~\cite{MWZ:2011}.

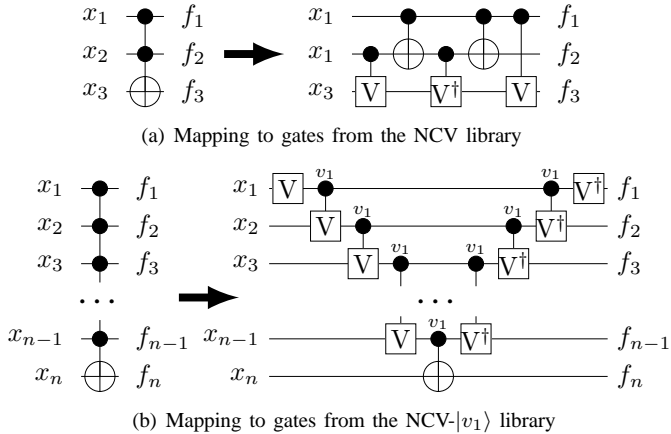
\begin{figure}[t]
\centering

\subfigure[Mapping to gates from the NCV library\label{fig:gate_mapping_ncv}]{\begin{tikzpicture}

\begin{scope}
\draw[line width=0.300000] (0.500000,1.250000) -- (1.000000,1.250000);
\draw (0.400000,1.250000) node [left] {$x_1$};
\draw (1.100000,1.250000) node [right] {$f_1$};
\draw[line width=0.300000] (0.500000,0.750000) -- (1.000000,0.750000);
\draw (0.400000,0.750000) node [left] {$x_2$};
\draw (1.100000,0.750000) node [right] {$f_2$};
\draw[line width=0.300000] (0.500000,0.250000) -- (1.000000,0.250000);
\draw (0.400000,0.250000) node [left] {$x_3$};
\draw (1.100000,0.250000) node [right] {$f_3$};
\draw[line width=0.300000] (0.750000,0.250000) -- (0.750000,1.250000);
\draw[fill] (0.750000,1.250000) circle (0.100000);
\draw[fill] (0.750000,0.750000) circle (0.100000);
\draw[line width=0.3] (0.75,0.25) circle (0.2) (0.75,0.05) -- (0.75,0.45);
\end{scope}

\draw[-latex,line width=3] (1.8,.75) -- ++(.8,0);

\begin{scope}[xshift=3cm]
\draw[line width=0.300000] (0.500000,1.250000) -- (3.000000,1.250000);
\draw (0.400000,1.250000) node [left] {$x_1$};
\draw (3.100000,1.250000) node [right] {$f_1$};
\draw[line width=0.300000] (0.500000,0.750000) -- (3.000000,0.750000);
\draw (0.400000,0.750000) node [left] {$x_1$};
\draw (3.100000,0.750000) node [right] {$f_2$};
\draw[line width=0.300000] (0.500000,0.250000) -- (3.000000,0.250000);
\draw (0.400000,0.250000) node [left] {$x_3$};
\draw (3.100000,0.250000) node [right] {$f_3$};
\draw[line width=0.300000] (0.750000,0.250000) -- (0.750000,0.750000);
\draw[fill] (0.750000,0.750000) circle (0.100000);
\draw[line width=0.3,fill=white] (0.55,0.05) rectangle ++(0.4,0.4); \node at (0.75,0.25) {\sf $\V$};
\draw[line width=0.300000] (1.250000,0.750000) -- (1.250000,1.250000);
\draw[fill] (1.250000,1.250000) circle (0.100000);
\draw[line width=0.3] (1.25,0.75) circle (0.2) (1.25,0.55) -- (1.25,0.95);
\draw[line width=0.300000] (1.750000,0.250000) -- (1.750000,0.750000);
\draw[fill] (1.750000,0.750000) circle (0.100000);
\draw[line width=0.3,fill=white] (1.55,0.05) rectangle ++(0.4,0.4); \node at (1.75,0.25) {\sf $\Vp$};
\draw[line width=0.300000] (2.250000,0.750000) -- (2.250000,1.250000);
\draw[fill] (2.250000,1.250000) circle (0.100000);
\draw[line width=0.3] (2.25,0.75) circle (0.2) (2.25,0.55) -- (2.25,0.95);
\draw[line width=0.300000] (2.750000,0.250000) -- (2.750000,1.250000);
\draw[fill] (2.750000,1.250000) circle (0.100000);
\draw[line width=0.3,fill=white] (2.55,0.05) rectangle ++(0.4,0.4); \node at (2.75,0.25) {\sf $\V$};
\end{scope}

\end{tikzpicture}}
\subfigure[Mapping to gates from the NCV-$|v_1\rangle$ library\label{fig:gate_mapping_ncvv1}]{\begin{tikzpicture}
\begin{scope}
\draw[line width=0.300000] (0.500000,2.250000) -- (1.000000,2.250000);
\draw (0.400000,2.250000) node [left] {$x_1$};
\draw (1.100000,2.250000) node [right] {$f_1$};
\draw[line width=0.300000] (0.500000,1.750000) -- (1.000000,1.750000);
\draw (0.400000,1.750000) node [left] {$x_2$};
\draw (1.100000,1.750000) node [right] {$f_2$};
\draw[line width=0.300000] (0.500000,1.250000) -- (1.000000,1.250000);
\draw (0.400000,1.250000) node [left] {$x_3$};
\draw (1.100000,1.250000) node [right] {$f_3$};

\draw[line width=0.300000] (0.750000,1) -- (0.750000,2.250000); 
\draw[fill] (0.750000,2.250000) circle (0.100000);
\draw[fill] (0.750000,1.750000) circle (0.100000);
\draw[fill] (0.750000,1.250000) circle (0.100000);

\begin{scope}[yshift=-.5cm]
\node[anchor=south] at (.75,1.1) {\Large \dots};

\draw[line width=0.300000] (0.500000,0.750000) -- (1.000000,0.750000);
\draw (0.400000,0.750000) node [left] {$x_{n-1}$};
\draw (1.100000,0.750000) node [right] {$f_{n-1}$};
\draw[line width=0.300000] (0.500000,0.250000) -- (1.000000,0.250000);
\draw (0.400000,0.250000) node [left] {$x_n$};
\draw (1.100000,0.250000) node [right] {$f_n$};

\draw[line width=0.300000] (0.750000,0.250000) -- (0.750000,1);
\draw[fill] (0.750000,0.750000) circle (0.100000);
\draw[line width=0.3] (0.75,0.25) circle (0.2) (0.75,0.05) -- (0.75,0.45);
\end{scope}
\end{scope}

\draw[-latex,line width=3] (1.8,0.8) -- ++(.8,0);

\begin{scope}[xshift=3cm]
\draw[line width=0.300000] (0.000000,2.250000) -- (4.500000,2.250000);
\draw (0.000000,2.250000) node [left] {$x_1$};
\draw (4.500000,2.250000) node [right] {$f_1$};
\draw[line width=0.300000] (0.000000,1.750000) -- (4.500000,1.750000);
\draw (0.000000,1.750000) node [left] {$x_2$};
\draw (4.500000,1.750000) node [right] {$f_2$};
\draw[line width=0.300000] (0.000000,1.250000) -- (4.500000,1.250000);
\draw (0.000000,1.250000) node [left] {$x_3$};
\draw (4.500000,1.250000) node [right] {$f_3$};

\draw[line width=0.300000] (0.750000,1.750000) -- (0.750000,2.250000);
\draw[line width=0.3,fill=white] (0.05,2.05) rectangle ++(0.4,0.4); \node at (0.25,2.25) {\sf $\V$};

\draw[fill] (0.750000,2.250000) circle (0.100000);
\node[anchor=south] at (0.750000,2.250000) {\scriptsize $v_1$};
\draw[line width=0.3,fill=white] (0.55,1.55) rectangle ++(0.4,0.4); \node at (0.75,1.75) {\sf $\V$};
\draw[line width=0.300000] (1.250000,1.250000) -- (1.250000,1.750000);
\draw[fill] (1.250000,1.750000) circle (0.100000);
\node[anchor=south] at (1.250000,1.750000) {\scriptsize $v_1$};
\draw[line width=0.3,fill=white] (1.05,1.05) rectangle ++(0.4,0.4); \node at (1.25,1.25) {\sf $\V$};

\draw[line width=0.300000] (1.750000,1.250000) -- ++(0,-.3);
\draw[fill] (1.750000,1.250000) circle (0.100000);
\node[anchor=south] at (1.750000,1.250000) {\scriptsize $v_1$};

\begin{scope}[yshift=-.5cm]
\node[anchor=south] at (2.250000,1.1) {\Large \dots};

\draw[line width=0.300000] (0.000000,0.750000) -- (4.500000,0.750000);
\draw (0.000000,0.750000) node [left] {$x_{n-1}$};
\draw (4.500000,0.750000) node [right] {$f_{n-1}$};
\draw[line width=0.300000] (0.000000,0.250000) -- (4.500000,0.250000);
\draw (0.000000,0.250000) node [left] {$x_n$};
\draw (4.500000,0.250000) node [right] {$f_n$};

\draw[line width=0.300000] (1.750000,0.750000) -- ++(0,.3);
\draw[line width=0.3,fill=white] (1.55,0.55) rectangle ++(0.4,0.4); \node at (1.75,0.75) {\sf $\V$};

\draw[line width=0.300000] (2.250000,0.250000) -- (2.250000,0.750000);
\draw[fill] (2.250000,0.750000) circle (0.100000);
\node[anchor=south] at (2.250000,0.750000) {\scriptsize $v_1$};
\draw[line width=0.3] (2.25,0.25) circle (0.2) (2.25,0.05) -- (2.25,0.45);

\draw[line width=0.300000] (2.750000,0.750000) -- ++(0,.3);
\draw[line width=0.3,fill=white] (2.55,0.55) rectangle ++(0.4,0.4); \node at (2.75,0.75) {\sf $\Vp$};

\end{scope}

\draw[line width=0.300000] (2.750000,1.250000) -- ++(0,-.3);
\draw[fill] (2.750000,1.250000) circle (0.100000);
\node[anchor=south] at (2.750000,1.250000) {\scriptsize $v_1$};

\draw[line width=0.300000] (3.250000,1.250000) -- (3.250000,1.750000);
\draw[fill] (3.250000,1.750000) circle (0.100000);
\node[anchor=south] at (3.250000,1.750000) {\scriptsize $v_1$};
\draw[line width=0.3,fill=white] (3.05,1.05) rectangle ++(0.4,0.4); \node at (3.25,1.25) {\sf $\Vp$};
\draw[line width=0.300000] (3.750000,1.750000) -- (3.750000,2.250000);
\draw[fill] (3.750000,2.250000) circle (0.100000);
\node[anchor=south] at (3.750000,2.250000) {\scriptsize $v_1$};
\draw[line width=0.3,fill=white] (3.55,1.55) rectangle ++(0.4,0.4); \node at (3.75,1.75) {\sf $\Vp$};

\draw[line width=0.3,fill=white] (4.05,2.05) rectangle ++(0.4,0.4); \node at (4.25,2.25) {\sf $\Vp$};

\end{scope}
\end{tikzpicture}
}
\caption{Mapping reversible circuits to quantum circuits}\label{fig:gate_mapping}
\end{figure}

\begin{example}
Consider a Toffoli gate with two control lines as shown in the left-hand side of
Fig.~\ref{fig:gate_mapping_ncv}.
A functionally equivalent realization in terms of gates from the NCV library is depicted in
the right-hand side of Fig.~\ref{fig:gate_mapping_ncv}.
\end{example}

Similar mappings exist for Toffoli gates with more than two control lines.
But with increasing number of control lines, the resulting quantum circuits become
more expensive, i.e.~require more quantum gates.
Furthermore, also the number of the \emph{ancillarly lines}, i.e.~the number of circuit lines which neither are a control line nor a target line, affect the size of the resulting quantum circuit. To provide some numbers,
Table~\ref{tab:barenco_costs} lists the respective costs for different Toffoli gate configurations according to the current state-of-the-art NCV mapping scheme introduced in~\cite{MWZ:2011}.

\subsection{Mapping to Gates from the NCV-$|v_1\rangle$ Library}\label{sec:mapping_ncvv1}

With the NCV-$|v_1\rangle$ library, also a corresponding mapping scheme has been 
introduced in~\cite{SWM:2012}.
This scheme fully exploits the $|v_1\rangle$-sensitivity of the control lines which 
enables a more efficient mapping of reversible gates than the mapping to gates from the 
NCV library.
The general principle of this mapping is illustrated by means of the following example.

\begin{example}
Consider a Toffoli gate with an arbitrary number of control lines as shown in the left-hand side of
Fig.~\ref{fig:gate_mapping_ncvv1}.
A functionally equivalent realization in terms of gates from the NCV-$|v_1\rangle$ is depicted in
the right-hand side of Fig.~\ref{fig:gate_mapping_ncvv1}.
First, all control lines are $|v_1\rangle$-sensitized, i.e.~($v_1$-controlled) $\V$-gates are applied
setting the values of the control lines to~$v_1$ iff they all have initially been set to~1.
By this, a $v_1$-controlled NOT gate ensures that the value of the target line is only flipped
iff all control lines have been set to~1. Afterwards, ($v_1$-controlled) $\Vp$-gates are applied
to de-sensitize the control lines.
\end{example}

Using this scheme, every Toffoli gate $T(C,x_j)$ with $x_j\in X$ and $C\subset X\backslash\{x_j\}$ can 
be mapped to an equivalent cascade of~$2\cdot|C|+1$ NCV-$|v_1\rangle$ quantum gates~\cite{SWM:2012}. 
In comparison to the mapping to gates from the NCV library, 
this is 
(1)~significantly more compact and (2)~does not even require ancillary lines.
Table~\ref{tab:costs_gate} provides a more precise comparison between these two mappings.
Note that the physical costs of the respective gates may differ in the respective libraries. However,
comparing the number of elementary gates seem to be an acceptable abstraction until the physically realizations eventually advanced.

\begin{table}[!t]
\centering
  \caption{Quantum costs}\label{tab:costs_gate}
{\setlength{\tabcolsep}{4pt}
\subfigure[Using NCV library\label{tab:barenco_costs}]{\begin{tabular}{|lc||r|r|r|r|r|r|}
\hline
      & &\multicolumn{6}{c|}{{Number of Ancillary Lines}} \\
& & 1 & 2 & 3 & 4 & 5 & 6\\
\hline
\hline
\multirow{13}{0cm}{\begin{turn}{90}{ Number of Control Lines}\end{turn}}&1 & 1 &  &  &  &  & \\
\cline{2-8}
&2 & 5 &  &  &  &  & \\
\cline{2-8}
&3 & 14 &  &  &  &  & \\
\cline{2-8}
&4 & 20 &  &  &  &  & \\
\cline{2-8}
&5 &32 &  &  &  &  & \\
\cline{2-8}
&6 & 44 &  &  &  &  & \\
\cline{2-8}
&7 &  64 & 56 &  &  &  & \\
\cline{2-8}
&8 &76 & 68 &  &  &  & \\
\cline{2-8}
&9 & 96 & 88 & 80 &  &  & \\
\cline{2-8}
&10 & 108 & 100 & 92 &  &  & \\
\cline{2-8}
&11 &  132 & 120 & 112 & 104 &  & \\
\cline{2-8}
&12 &  156 & 132 & 124 & 116 &  & \\
\cline{2-8}
&13 &  180 & 156 & 148 & 136 & 128 & \\
\cline{2-8}
&14 &  204 & 180 & 172 & 148 & 140 & \\
\cline{2-8}
&15 &  228 & 204 & 198 & 172 & 160 & 152\\
\hline
\end{tabular}}
\subfigure[Using NCV-$|v_1\rangle$ lib.\label{tab:new_costs}]{\begin{tabular}{|cc||r|r|}
\hline
& &  &   \\
& & Costs & $\Delta$ \\
\hline
\hline
\multirow{13}{0cm}{\begin{turn}{90}{ Number of Control Lines}\end{turn}}&1 & 1 & 0\%  \\
\cline{2-4}
&2 & 3 & 40\%  \\
\cline{2-4}
&3 & 5 & 64\%  \\
\cline{2-4}
&4 &7 &  65\% \\
\cline{2-4}
&5 &9 &  72\%\\
\cline{2-4}
&6 & 11 & 75\%\\
\cline{2-4}
&7 &  13 &  77-80\%\\
\cline{2-4}
&8 &15 & 78-80\%\\
\cline{2-4}
&9 & 17 & 79-82\%\\
\cline{2-4}
&10 & 19 &  79-82\%\\
\cline{2-4}
&11 &  21 & 80-84\%\\
\cline{2-4}
&12 &  23 &  80-85\%\\
\cline{2-4}
&13 &  25 &  80-86\%\\
\cline{2-4}
&14 &  27 &  81-87\%\\
\cline{2-4}
&15 &  29 & 81-87\%\\
\hline
\end{tabular}}
}
\end{table}

\section{Conclusions and Future Work}

In this technical report, the ideas and concepts of the 
\mbox{NCV-$|v_1\rangle$} library have been described in more detail. 
Note thereby that this work as well as the previous contributions in~\cite{MS:00,SWM:2012} 
just provided a theoretical and conceptual discussion on the applicability of the
NCV-$|v_1\rangle$ library. Physical realizations and concepts on how to realize this
library still is subject to future work. This includes (1)~direct realizations of the qudits, (2)~the emulation of qudits e.g.~by existing qubit-realizations,
and (3)~the compatibility to existing fault-tolerant quantum error correction protocols .
Accordingly also questions on the implementation costs of the respective gates are left for future work. Thus far and as common for other gate libraries, we just applied the number of elementary quantum gates as optimization criterion.

\bibliographystyle{IEEEtran}
\bibliography{lit_header,lit_mymisc,lit_myrev,lit_others,lit_othersrev,myloc} 

\end{document}